\documentclass[aip, jcp, reprint]{revtex4-1}
\usepackage{amsmath}
\usepackage{mathtools}
\usepackage{graphicx}
\usepackage{dcolumn}
\usepackage{bm}

\begin{document}

\preprint{AIP/123-QED}

\title{Insights into DNA-mediated interparticle interactions from a coarse-grained model}

\author{Yajun Ding}

\author{Jeetain Mittal}
\email{jeetain@lehigh.edu}
\affiliation{Department of Chemical and Biomolecular Engineering, Lehigh University, Bethlehem, Pennsylvania 18015, USA\\}

\date{\today}

\begin{abstract} 

DNA-functionalized particles have great potential for the design of complex 
self-assembled materials. The major hurdle in realizing crystal structures from  
DNA-functionalized particles is expected to be kinetic barriers that trap 
the system in metastable amorphous states. Therefore, it is vital to explore the 
molecular details of particle assembly processes in order to understand the underlying 
mechanisms. Molecular simulations based on coarse-grained models can provide 
a convenient route to explore these details. Most of the currently available 
coarse-grained models of DNA-functionalized particles ignore key chemical and structural 
details of DNA behavior. These models therefore are limited 
in scope for studying experimental phenomena. 
In this paper, we present a new coarse-grained
model of DNA-functionalized particles which incorporates some of the desired features of DNA behavior. 
The coarse-grained DNA model used here provides explicit DNA representation 
(at the nucleotide level) and complementary interactions between Watson-Crick 
base pairs, which lead to the formation of single-stranded hairpin and double-stranded 
DNA. Aggregation between multiple complementary strands is also prevented in our model. 
We study interactions between two DNA-functionalized 
particles as a function of DNA grafting density, lengths of the hybridizing and non-hybridizing parts 
of DNA, 
and temperature. 
The calculated free energies as a function of pair distance between particles qualitatively resemble 
experimental measurements of DNA-mediated pair interactions.

\end{abstract}

\maketitle

\section{Introduction}

It is now well recognized that self-assembly of particles functionalized with biomolecules is a promising way 
to form unique nanostructured materials~\cite{Geerts2010, Michele2013}. The distinct advantage of inter-particle interactions mediated by 
DNA molecules is the specificity of Watson-Crick (WC) pairing~\cite{Alivisatos1996, Mirkin1996}; the functionalized 
particles can be bridged together either by direct hybridization of  complementary single-stranded DNA (ssDNA) 
molecules (sticky end) or indirectly via ``linker" DNA molecules that can bind simultaneously to the complementary 
ssDNA sequences on two different particles\cite{Geerts2010}. 
While significant progress has been made in recent years, the fundamental details of  
DNA-mediated particle assembly are not very well understood~\cite{Geerts2010, Michele2013}. 
Specifically, the assembly of micron-sized particles into crystalline structures is still quite challenging, 
though nanoparticles have been assembled into a wide variety of periodic arrangements via 
DNA-mediated interactions~\cite{Park2008, Nykypanchuk2008, Macfarlane2011}. 


Computer simulations can provide a convenient route to explore the large parameter space of  
DNA-functionalized particles (DFPs), which can be useful for developing basic understanding and to test existing 
theoretical design models~\cite{Macfarlane2011,srinivasan2013designing, Mirjam2011, Mladek2012}. 
Using all-atom models to study the properties of DFPs is still beyond the current 
state-of-the-art computational capabilities~\cite{Lee2009a,Lee2009b,Ngo2012}. Simple and accurate 
coarse-grained (CG) models are, therefore, needed to overcome existing computational challenges and are 
actively being developed. 
Starr and co-workers \cite{Olivia2011, Vargas2011, Dai2010, Hsu2008, Hsu2010} have developed a ``two-bead'' DFP 
model in which each nucleotide is represented by two force sites, one for the phosphate-sugar backbone and one 
for the nitrogenous base. The backbone beads are connected by the standard finite extensible nonlinear elastic (FENE) 
bond potential. The beads representing the nucleotide bases are also connected to the backbone beads by the FENE 
potential.  To model WC pairing, the 12-6 Lennard-Jones (LJ) nonbonded potential between complementary bases is used. 
The particle itself is modeled as a single spherical or icosahedral core for simplicity. 
This model has extensively been applied to understand nanoparticle dimers, the stability of nanoparticle 
crystals, polymorphism, equilibrium clustering and dynamics~\cite{Olivia2011, Vargas2011, Dai2010, Hsu2008, Hsu2010}. 
Seifpour {\em et al.} also proposed a 
modified model based on this model and applied it to study the effect of DNA strand composition and 
sequence on the structure and thermodynamics of DFPs~\cite{Seifpour2013}.

Li {\em et al.} \cite{Ting2012, Ting2013} recently introduced a CG DFP model which is based on an 
earlier model by Travesset {\em et al.} \cite{Knorowski2011, Knorowski2012}. In this model, the  DNA molecule 
is made up of three parts: single-stranded DNA, double-stranded DNA (dsDNA), and a ``sticky end". 
Connected beads of different sizes are used to represent ssDNA and dsDNA. The sticky end is modeled by  
multiple beads to take into account the selectivity and directionality of hydrogen bonding between complementary 
DNA bases. With this model, Li {\em et al.} reproduced all nine crystal structures observed 
experimentally  
by Macfarlane et al. \cite{Macfarlane2011}, and also proposed new linker sequences 
for future experiments. 
Frenkel and co-workers \cite{Martinez2011, Martinez2010, Mirjam2011,  Mladek2012} have developed a 
CG ``core-blob" 
DFP model by further reducing the DNA degrees of freedom. In their model, the sticky end of tethered DNA is modeled as a 
blob connected to an effective spherical core. 
This model can capture several important features of DNA-mediated particle assembly, from pair-particle 
interactions~\cite{Mirjam2011} to thermodynamic phase transitions~\cite{Mladek2012, Martinez2010}. 

When moving from the CG model with explicit DNA-like chains to a simplified core-blob model, the structural and chemical details 
of DNA molecules are gradually lost. Although it is obvious that significant coarse graining is needed to explore 
DFP system behavior (in particular spontaneous crystallization), certain molecular details, which 
are often ignored in these previous models, are important to 
understand DNA-mediated particle assembly. For example, most current models use an average base 
representation (ignoring the differences between A:T and G:C pair interactions), thereby 
neglecting the dependence of particle interactions on the DNA sequence. Not only is the difference between strengths of 
G:C pairing (stronger due to the presence of three interbase hydrogen bonds) compared to A:T pairing (two interbase 
hydrogen bonds) important, but the actual DNA sequence  
can also affect properties of hybridized DNA molecules~\cite{Petr2012, Seifpour2013}. 
In addition, it has been reported that in order to capture surface-adsorbed or interfacial DNA structure, one needs to 
account for non-WC base pairing as well, since dsDNA melting and ssDNA properties will be affected by these 
additional interactions~\cite{Margaret2011, Shankar2012, Roxbury2012, Roxbury2013}. Base stacking interactions, which drive the coplanar 
alignment of neighboring bases, are also important to the overall stability of helical duplexes and are also 
sequence-dependent~\cite{Yakovchuk2006}. Furthermore, the intra-particle interactions, which 
are the interactions between DNA strands on the same particle, are often ignored but can be quite important for  
particle diffusion and kinetics of assembly~\cite{Chremos2011}. 

In this paper, we present a new CG DFP model in which we use a two-bead representation for each nucleotide of 
tethered DNA molecules. The DNA model, which is based on a previously proposed model by Dorfman and co-workers~\cite{Kenward2009}, 
displays typical experimentally observed temperature melting behavior for dsDNA hybridization and ssDNA hairpin formation. 
The DNA strands are connected to spherical particle cores (made up of smaller point particle beads) by FENE bonds to 
simulate DFPs.  
We use this model to study the potential of mean force (PMF) between two DFPs as a function of 
DNA grafting density, spacer length, sticky end length, and temperature.  
The results presented in this paper suggest that this new DFP model can be useful for 
studying the thermodynamics and kinetics of particle assembly mediated by DNA hybridization.

The paper is organized as follows: In section \MakeUppercase{\romannumeral 2},  we outline the details of the model 
and potential parameters. 
In section \MakeUppercase{\romannumeral 3}A, we present data on the melting behavior of ssDNA hairpin and dsDNA in the 
absence of particles. Next, in section \MakeUppercase{\romannumeral 3}B, we show potentials of mean force between 
two DFPs and discuss their connection with previous experimental and theoretical data. 
In section \MakeUppercase{\romannumeral 4}, we conclude with some key observations from this work and discuss our 
future goals.

\section{model and simulation details}

\subsection{DNA and particle models}

In order to achieve an accurate description of DNA behavior in a CG DFP 
model, the DNA model itself should be 
suitable for studying the properties of ssDNA and dsDNA. The model proposed 
here is expected to include the following 
key aspects of DNA behavior that may be important 
for understanding the assembly of DFPs: (i) attractive interactions between 
complementary base pairs A:T and G:C, 
(ii) attractive interactions between adjacent bases to capture base stacking, 
(iii) the ability to study sequence-dependent behavior, (iv) hairpin formation in 
a suitable ssDNA sequence with complementary WC pairs at both 
ends, 
(v) hybridization between two fully or partially complementary strands, and 
(vi) no multi-strand association or aggregation in the form of large bundles. 

\begin{figure}
\includegraphics[width= 0.95\columnwidth]{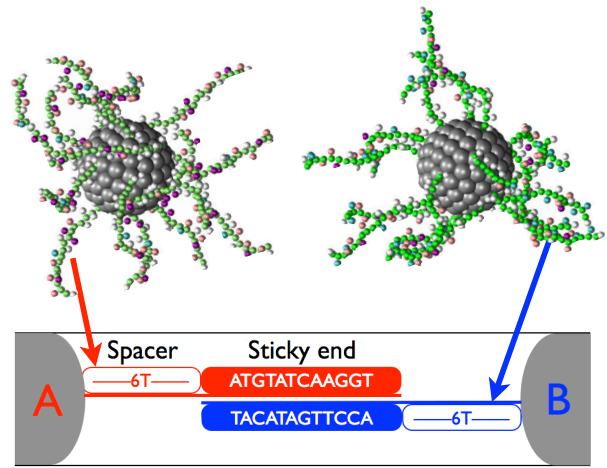}
\caption{Coarse-grained DNA functionalized particle model.  
Two particles with multiple DNA molecules attached to the particle surface atoms are shown. The 
DNA sequence is composed of two parts, the spacer (6T: TTTTTT) and the sticky end as shown in the 
schematic.}
\label{atomvscg}
\end{figure}

It is also desirable to include base pair dependent 
non-WC interactions~\cite{Shankar2012}, but is something that will require 
extensive parameterization and testing. We plan to do this in future 
refinement of the model proposed here.
By taking into account the requirements above, we have developed our CG model, 
which is based on an earlier 
DNA model proposed by Dorfman and co-workers~\cite{Kenward2009}. Our choice is based on a compromise between computational efficiency and accurate representation of DNA behavior. Other DNA models are available in the literature, which can be computationally more efficient or physically more accurate. In our model 
(see Figure~\ref{atomvscg}), a two bead representation is used for a nucleotide, one bead for the 
phosphate-sugar backbone and another for the nitrogenous base.

The connectivity between base and backbone beads and between adjacent backbone beads is modeled by 
the finite extensible nonlinear elastic (FENE) bond potential~\cite{Kremer1990}, which is given by
\begin{equation}
U_{\text{bond}}(r_{ij}) = -\frac{k_{\mathrm {bond}}}{2}R_{0}^{2}\text{ln}\left[1-\left(\frac{r_{ij}}{R_{0}}\right)^{2}\right],
\end{equation}
where $k_{\mathrm {bond}}$ is the effective strength of the potential, and $R_{0}$ is the cutoff distance at which this potential 
diverges. Following Kenward {\em et al.}, we use $k_{\mathrm {bond}} = 30\epsilon/\sigma^{2}$ and $R_{0}=1.5\sigma$ 
here ~\cite{Kenward2009}. Here, $\epsilon$ is the characteristic energy scale and 
$\sigma$ is 
the characteristic length scale. The angle potential is used to provide additional bending rigidity to the backbone with the following functional form:
\begin{equation}
U_{\text{angle}}(\theta) =\frac{k_{\mathrm {angle}}}{2}(cos\theta+cos\theta_{0})^{2},
\end{equation}
where $\theta$ the angle between three adjacent backbone beads, $\theta_{0} = 180^{\circ}$ is the reference bending angle, and $k_{\mathrm {angle}}=24\epsilon$ is the stiffness parameter.
Inter-chain and intra-chain stacking (st) and WC hydrogen bonding (hb) attractions between 
DNA base beads are modeled as~\cite{Kenward2009}
\begin{equation}
U_{k}(r_{ij})=-\epsilon u_{k}\delta_{ij}^{k}\left\lbrace \text{exp}\left(20\left[\frac{r_{ij}}{\sigma}-\Gamma_{s}\right]\right)+1\right\rbrace^{-1}, 
\end{equation}
where $k\in\{{\mathrm {hb}}, {\mathrm {st}}\}$ and $\Gamma_{s}$ sets the range of interaction; we use $\Gamma_{s}=1.5$. The interactions between possible base pairs out of the four base alphabet 
(A, T, C or G) can be simply represented in a matrix form $[\delta_{ij}^{k}]$ 
as~\cite{Kenward2009}
\begin{eqnarray}
[\delta_{ij}^{st}] =
\begin{pmatrix}
       \delta_{AA}^{st} & \delta_{AT}^{st} & \delta_{AC}^{st} & \delta_{AG}^{st}          \\[0.5em]
       \delta_{TA}^{st} & \delta_{TT}^{st} & \delta_{TC}^{st} & \delta_{TG}^{st}           \\[0.5em]
       \delta_{CA}^{st} & \delta_{CT}^{st} & \delta_{CC}^{st} & \delta_{CG}^{st}           \\[0.5em]
       \delta_{GA}^{st} & \delta_{GT}^{st} & \delta_{GC}^{st} & \delta_{GG}^{st}
\end{pmatrix}
                     =
\begin{pmatrix}
       \frac{3}{4} & \frac{1}{2} & \frac{1}{2} & \frac{3}{4}           \\[0.5em]
       \frac{1}{2} & \frac{1}{4} & \frac{1}{4} & \frac{1}{2}           \\[0.5em]
       \frac{1}{2} & \frac{1}{4} & \frac{3}{4} & \frac{3}{4}           \\[0.5em]
       \frac{3}{4} & \frac{1}{2} & \frac{3}{4} & 1
\end{pmatrix}, 
\end{eqnarray} 

\begin{eqnarray}
[\delta_{ij}^{hb}]=\begin{pmatrix}
       \delta_{AA}^{hb} & \delta_{AT}^{hb} & \delta_{AC}^{hb} & \delta_{AG}^{hb}           \\[0.5em]
       \delta_{TA}^{hb} & \delta_{TT}^{hb} & \delta_{TC}^{hb} & \delta_{TG}^{hb}           \\[0.5em]
       \delta_{CA}^{hb} & \delta_{CT}^{hb} & \delta_{CC}^{hb} & \delta_{CG}^{hb}           \\[0.5em]
       \delta_{GA}^{hb} & \delta_{GT}^{hb} & \delta_{GC}^{hb} & \delta_{GG}^{hb}
\end{pmatrix}
                     =
\begin{pmatrix}
       0 & \frac{2}{3} & 0 & 0           \\[0.1em]
       \frac{2}{3} & 0 & 0 & 0           \\[0.1em]
       0 & 0 & 0 & 1           \\[0.1em]
       0 & 0 & 1 & 0
\end{pmatrix}.
\end{eqnarray}

In Eq. (3), {$u_{k}$ sets the relative energy scale between base stacking $u_{st}$ and hydrogen 
bonding $u_{hb}$. We use $u_{st} = 2.5$ and $u_{hb} = 1.0$ and keep $u_{st}/u_{hb} = 2.5$ as suggested previously by Linak 
{\em et al.}~\cite{Linak2011}. 
The stacking potential $U_{\mathrm {st}}$ is only applied to adjacent bases in a DNA strand connected 
to backbone beads $i$ and $i\pm$1, which are less than distance $2\sigma$ apart. 
The hydrogen bonding potential $U_{\mathrm {hb}}$ is applied between all possible 
WC inter-strand base pairs and intra-strand base pairs, which are less than distance $2\sigma$ apart, 
excluding 
adjacent ($i$ and $i\pm$1) and next-nearest neighbors ($i$ and $i\pm$2).  
In addition, short-range repulsive  interactions between backbone-base and base-base beads are 
modeled by the Weeks-Chandler-Andersen (WCA) potential as
\begin{equation}
 U_{\text{WCA}}(r_{ij}) =
  \begin{dcases}
   4\epsilon\left[\left(\frac{\sigma}{r_{ij}}\right)^{12}-\left(\frac{\sigma}{r_{ij}}\right)^{6}\right]+\epsilon, & \text{if } r_{ij} \leq 2^{\frac{1}{6}}\sigma, \\
             0,& \text{if } r_{ij} > 2^{\frac{1}{6}}\sigma.
  \end{dcases}
\end{equation}

The repulsive interactions between negatively charged backbone beads are modeled by the Yukawa 
potential (screened electrostatics) as~\cite{Kremer1987} 
\begin{eqnarray}
U_{ele}(r_{ij}) &=& A\frac{e^{-\kappa r_{ij}}}{r_{ij}}, \text{ if } r_{ij} \le r_{\mathrm {cut}}, \\ \nonumber
&=& 0, \text{ if } r_{ij} > r_{\mathrm{cut}},
\end{eqnarray}
where $A$ is a prefactor with units of energy$\times$distance, $\kappa=2/\sigma$ is 
the screening length, and $r_{\mathrm {cut}}=3.2 \sigma$ is the potential cutoff 
value. 
In our reduced unit model, the actual strength of electrostatic interactions 
(A = 100 $\epsilon\sigma$) is selected so as to prevent 
triple- or multi-strand aggregation as shown in Figure~\ref{cluster}. 
The undesired aggregation of multiple DNA strands results from the absence of electrostatic repulsion between 
backbone beads, whereas well-separated dsDNA pairs form in the presence of 
such repulsion modeled as the Yukawa interaction potential (eq. 7). We note that 
aggregation of multiple DNA strands can also be avoided by accounting for the  
directional nature of hydrogen bonding interactions between DNA base beads with 
a multi-body interaction potential instead of the isotropic distance dependent 
potential given by eq. 3. 
To account for electrostatic interactions in a more 
transparent manner, we are currently developing a DNA model in real units with appropriate size scaling for 
the backbone and base beads as well as disparate bond lengths for backbone-backbone and base-backbone bonds~\cite{Boyer2014}. 
No attempt is made here to convert the reduced simulation units to SI units as was done previously by Linak et al.~\cite{Linak2011, Linak2010} for the temperature based on the experimental data. Without significant experimental input, such a mapping will likely yield inconsistent results.

\begin{figure}
\includegraphics[width= \columnwidth]{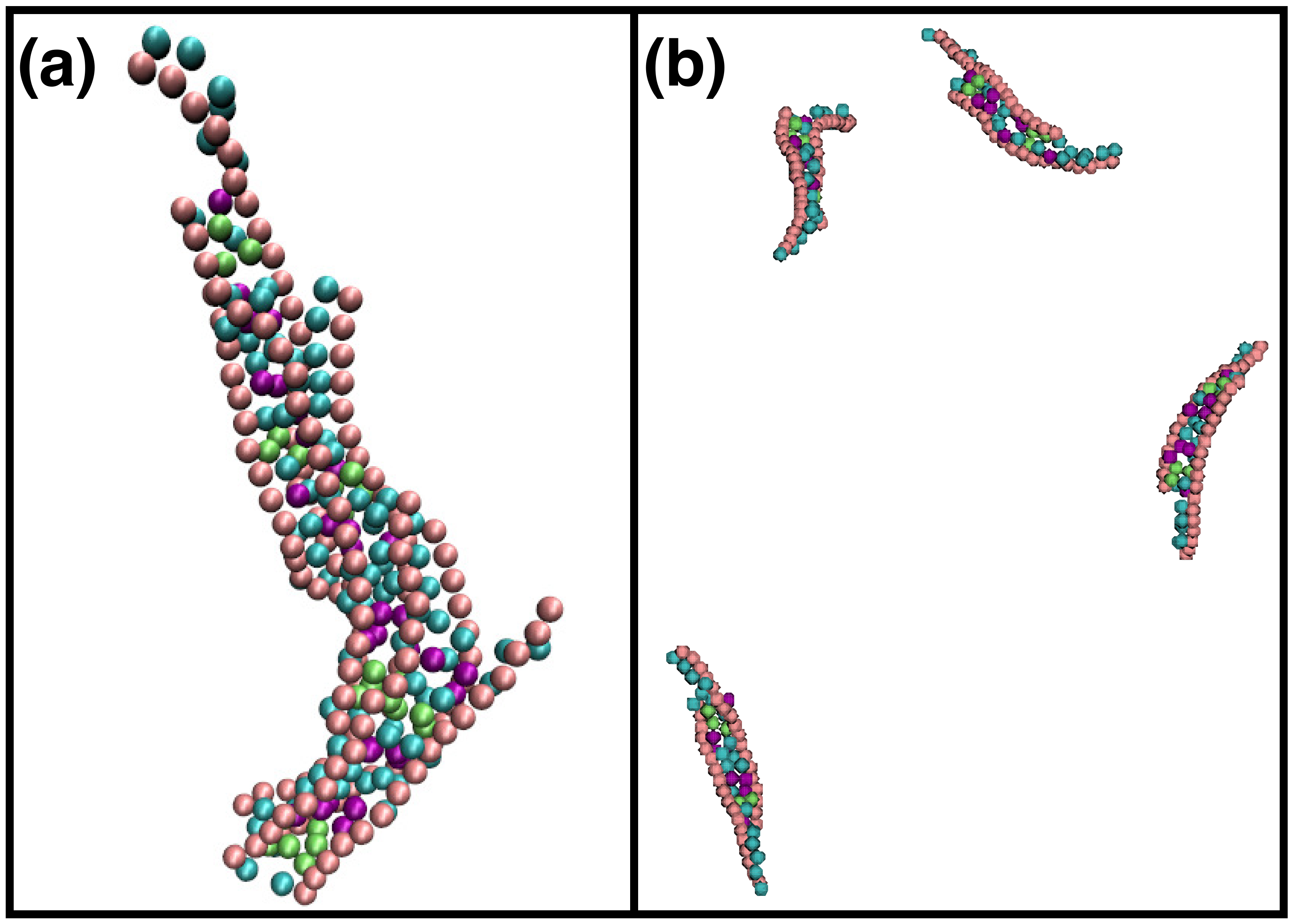}
\caption{Simulation snapshots of four pairs of partially complementary DNA 
sequence (see Figure~\ref{atomvscg}) without (a) and with (b) 
the Yukawa interaction potential temperature below the DNA melting temperature. 
}
\label{cluster}
\end{figure}

The particles are modeled as rigid body hollow spheres (radius $5\sigma$ used throughout this work),  
and each particle is made up of 100 beads uniformly distributed on the surface of 
a sphere~\cite{Simon2013}. The particle beads are uncharged and interact with DNA and other particle  
beads via the WCA 
interaction potential. 
Single strand DNA molecules are covalently linked to the particle surface beads by the FENE bond 
potential. We select particle surface beads that are farther than a certain distance value, which varies with DNA grafting density, to ensure uniform distribution of DNA strands on the particle surface. 
Figure~\ref{atomvscg} shows a snapshot of two CG DFPs with 
several ssDNA molecules attached at random locations to the particle surface. 
In our model, DNA molecules that are
attached to the same particle can interact via the WC base pair interactions, in contrast to previous 
CG DFP models that neglect intra-particle base pair interactions even between A:T and G:C base 
pairs. Therefore, appropriate DNA sequences in our model can 
form hairpin or loop structures, which have been proposed as ideal candidates for novel 
design schemes~\cite{Leunissen2009a, Leunissen2009b, Leunissen2010, Angioletti2012}.  
Both inter-particle and intra-particle base pair interactions are naturally incorporated in 
our CG DFP model.

\subsection{Simulation details}

To enhance equilibrium sampling, the replica exchange molecular dynamics (REMD) method 
is used for all of our simulations~\cite{Sugita:1999p8494}. REMD simulations are carried out under 
the canonical ensemble using a Langevin thermostat with damping parameter $\tau$ = 1 $(\epsilon/m/\sigma^{2})^{-1/2}$ in a cubic simulation box with periodic boundary conditions in all directions. The size of the simulation box is chosen to be $100\sigma$ so that there are 
no interactions between molecules and their periodic images. In our REMD simulations, we  
typically use 16 to 48 temperature replicas to obtain sufficient exchange probability and, 
therefore, convergence. The replica temperatures were chosen by trial and error such that the potential energy distributions adjacent replicas overlap sufficiently to ensure exchange between any two adjacent replicas with at least 50\% probability. The total number of timesteps that we performed for each simulation is of the order of $10^{8}$ 
steps with a step size of $\Delta t$ = 0.01 $(\epsilon/m/\sigma^{2})^{-1/2}$. 
About $10^{8}$ timesteps per replica are required to obtain a reasonable estimate of the melting profiles of the hairpin and the duplex, which takes approximately 144 hours of processor time (for 32 replicas) on the Lehigh computing server corona (corona.cc.lehigh.edu) equipped with AMD Opteron 8-core 6128, 2GHz processors. For DFPs, we need about $5\times10^{8}$ timesteps per replica, requiring approximately 2304 hours of processor time (for 32 replicas) on the Lehigh computing server corona. 

The DNA strands (hairpin or ssDNA) or DFPs are placed randomly inside the simulation box. The initial part of the simulation is discarded as equilibration, and the equilibration length is decided based on observables such as number of hydrogen bonds and distance between the particles. For hairpin and dsDNA duplex simulations the equilibration length is less than $10^{7}$ steps, while the equilibration length for DFPs is about $10^{8}$ steps. We use the block averaging scheme to estimate errors by dividing the production data into three sets and calculating the standard deviation from the mean. In case of the duplex/hairpin, the error bars of melting curves are smaller than the symbol size. 

When simulating DFPs, we also tested to ensure that results remain unchanged with a box size of 200$\sigma$. For the larger box size, we find it useful to employ umbrella sampling simulation~\cite{torrie1977} to obtain free energies as a function of pair distance between particles in a reasonable time. This also helps validate results obtained from REMD simulation. For umbrella sampling, we use harmonic umbrella potential with spring constant = 5.0 $\epsilon/\sigma^{2}$ along the pair distance between the particles with a sufficient number of copies to obtain overlap between distance distributions.

Finally, we have used two different pairs of complementary DNA sequences in our simulation~\cite{Rogers2011, Biancaniello2007}: a 12-mer, long sticky 
end (TTTTTTATGTATCAAGGT and TTTTTTACCTTGATACAT) and a 7-mer, short sticky end 
(TTTTTTTTTTTGTCTACC and TTTTTTTTTTTGGTAGAC). 

\section{results}
\subsection{Melting behavior of ssDNA hairpin and dsDNA}

\begin{figure}
\includegraphics[width= \columnwidth]{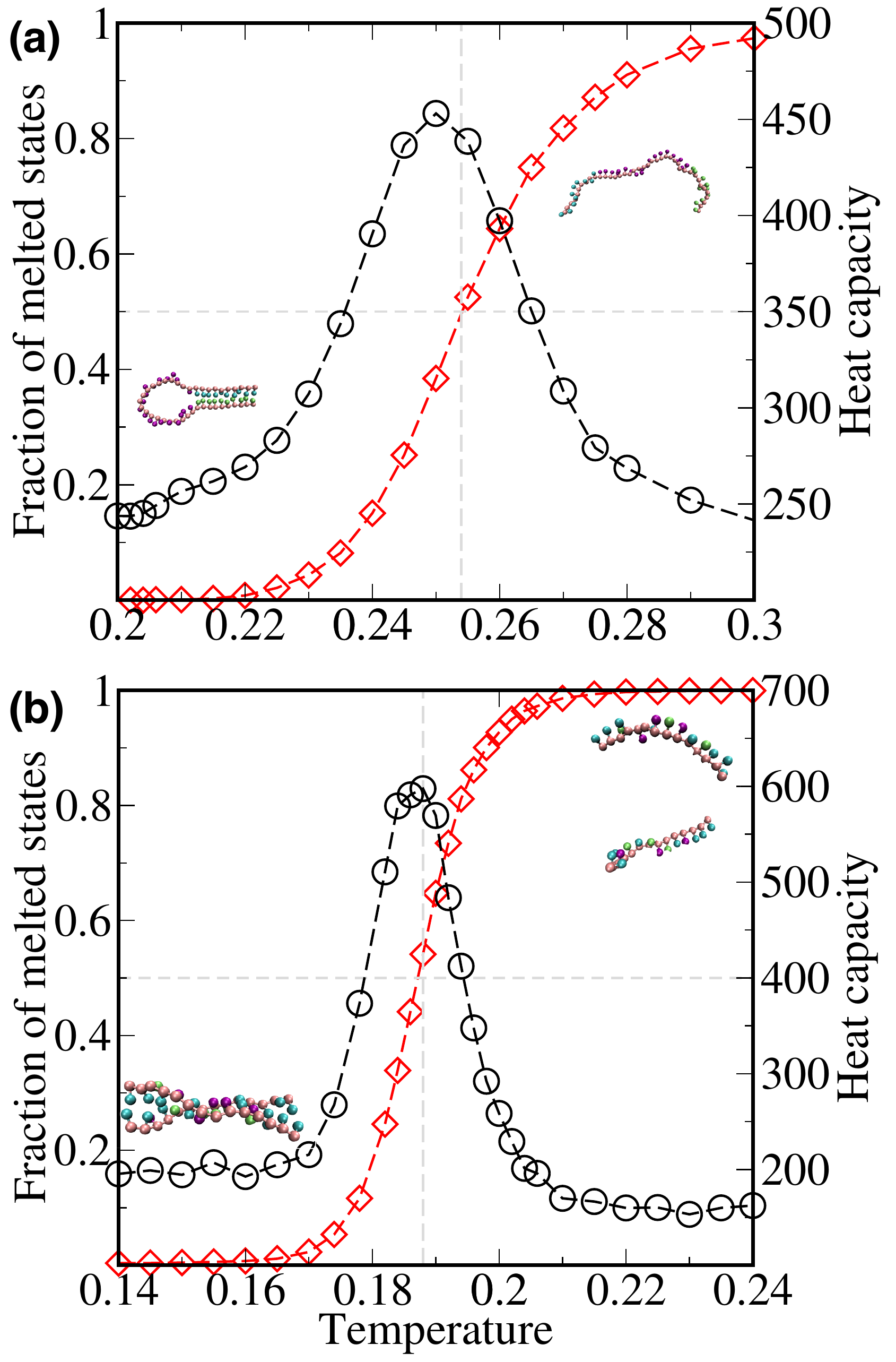}
\caption{Thermodynamic melting behavior. The fraction of melted states as a 
function of temperature (red symbols) (and typical DNA configurations) and heat capacity (black symbols) are shown for (a) a ssDNA hairpin:$\text{A}_{10}\text{G}_{20}\text{T}_{10}$ and (b) a pair of dsDNA  S1S2:GCGTCATACAGTGC. 
}
\label{meltdna}
\end{figure}

In order to validate the CG DNA model presented in section IIA, we first present the simulation data to characterize the melting behavior of ssDNA and dsDNA. The melting temperature is identified as the temperature at which the heat capacity (calculated from potential energy fluctuations) shows a maximum, the so-called calorimetric definition~\cite{Kim2008}. We also use a structure based definition to define melted or unhybridized states. We define ssDNA hairpin or dsDNA duplex configurations to be hybridized (see Supplemental Material (SM) Figure S1~\cite{sm}) if at least a certain number of the possible complementary base pairs are bonded. Up to a certain threshold value of number of bonded pairs (see SM Figures S2 and S3 ~\cite{sm}), the melting curves are quite similar. The melting temperature, identified from this structure based criteria as the temperature at which 50\% of the states are melted, is similar to the calorimetric definition based on the heat capacity.

Figure~\ref{meltdna} shows the melting curves for a ssDNA 
hairpin: $\text{A}_{10}\text{G}_{20}\text{T}_{10}$ and 
a dsDNA S1S2: GCGTCATACAGTGC obtained from REMD simulations. The shape of these melting 
curves is similar to what is expected based on thermodynamic melting behavior of 
hairpins and dsDNA formation from experiments~\cite{Wallace2001,Owczarzy2004}, although the widths of the melting transitions are much broader as found by Linak et al.~\cite{Linak2010, Linak2011} as well for a similar model as used here. Such broad transitions (unphysical) are also found in more detailed atomistic models of biomolecular systems such as proteins~\cite{Best2010}. 
From Figure~\ref{meltdna} we observe that at very low temperatures almost all states are found to be hybridized and at high temperatures all states are melted. The 
transition between these two states is occurring in a cooperative manner, more 
so in the case of dsDNA. The temperature at which the fraction of melted states equals  
the fraction of hybridized states can be defined as the melting temperature. 
We note that the melting curve obtained from simulation of a single pair of duplex in the canonical ensemble is different from the bulk experimental situation involving hybridization between many pairs of complementary strands, due to finite-size effects, as discussed in detail by Ouldridge et al.~\cite{Ouldridge2012, Ouldridge2010} If a comparison between simulation data from a finite system and experiment is desirable, melting temperature can be estimated as the temperature at which 33\% of states are melted, as opposed to 50\%. 
We also note that in the case of dsDNA hybridization, the melting temperature is 
also a function of the DNA concentration or simulation box volume for a fixed number of 
DNA strands.

Based on the results above, the CG DNA model is expected to capture the essential thermodynamics of dsDNA hybridization and ssDNA properties. Specifically, cooperative transitions between hybridized and unhybridized states for dsDNA and ssDNA, as a function of temperature are captured. Next, we use this model to study 
interactions between a pair of DNA-functionalized particles. 

\subsection{Pair interactions between DNA-functionalized particles}

Since attentive control of particle interactions is needed to obtain  
ordered structures, several previous efforts have been focused on understanding 
pair interactions between DNA-functionalized particles~\cite{}.  
On the experimental side, Crocker and co-workers measured DNA-mediated 
pair interactions between micron-sized particles.~\cite{biancaniello2005,Rogers2011} 
The separation distances between particles were obtained by confining two DFPs in  
harmonic potentials in an optical tweezer setup. The pair interaction 
potential (or free energy as a function of distance) is then obtained by the Boltzmann 
relation, accounting for optical forces. These direct measurements have allowed 
the development of statistical physics-based theoretical models for understanding the 
underlying physical phenomena~\cite{Rogers2011} of interactions between micron-sized 
particles. Though experimentally very challenging, it is desirable to obtain such 
information for interactions between nanoparticles as well.

Estimates of pairwise interactions between nanoparticles have been obtained by simulation 
with the help of CG DFP models (see section I)~\cite{Largo2007, Panagiotis2013}. 
These previous studies consider pair interactions between DFPs for very 
low DNA grafting densities of 4 to 6 DNA molecules per particle. 
Frenkel and co-workers~\cite{Mladek2012,Patrick2012} have developed numerical 
schemes to extract pair interactions using their core-blob DFP model with a 
combination of statistical mechanics theory and Monte Carlo simulations.  
Our new DFP model presented here is meant to act as a bridge between the core-blob 
type simplified description (which is computationally tractable even 
for phase behavior calculations) and experiment by including a more detailed description 
of DNA degrees of freedom (see section IIA). Ultimately, it may turn out that some of these 
details are not necessary to capture essential details of DFP assembly thermodynamics,  
including phase diagrams. But we anticipate that some of these additional details in our 
model may be important for understanding issues related to particle dynamics and the kinetics 
of assembly. 

\begin{figure}[b]
\includegraphics[width= 0.9\columnwidth]{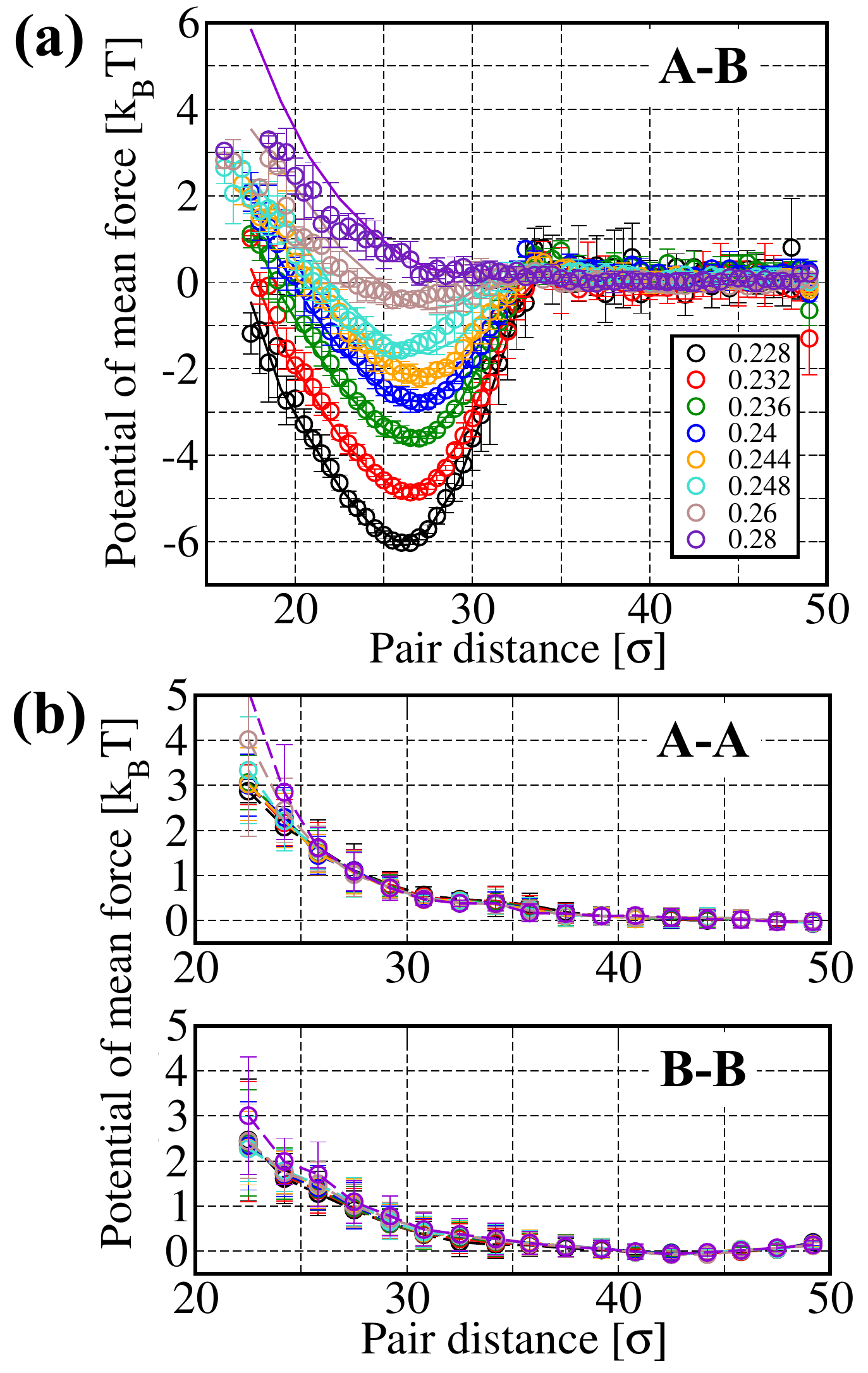}
\caption{Potential of mean force (PMF) calculated for a grafting density of 16 DNA 
molecules per particle with long sticky end for pairs A-B (a), A-A (b, top panel), 
and B-B (b, bottom panel). The data from simple binning at a single temperature 
shown by empty symbols, and solid lines are estimates from the 
weighted histogram analysis method. Symbol and line colors are used to  
distinguish between various temperatures as labeled in panel a.}
\label{pmfcb}
\end{figure}

We simulate two DFPs grafted with either the 12-mer long sticky end or the 7-mer short sticky end as described previously. 
The potential of mean force (PMF) or free energy as a function of pair distance 
between two DFPs, with respect to unhybridized particles, is calculated as 
\begin{equation}
\Delta F_{I}(r) = -k_{B}T\text{ln}[g(r)] + c,
\end{equation}
where $g(r)$ is the pair correlation function (normalized with respect to an 
ideal gas), $k_{B}$ is the Boltzmann constant, c is an additive constant, 
and $I$ is used to identify particle types in a pair, i.e., $I =\{\text{A-A},\text{B-B}, \text{A-B}\}$. In addition to binning data collected at a single temperature in an REMD simulation, 
we also use the weighted histogram analysis method 
(WHAM)~\cite{Kumar1992,Shirts2008} to combine data from all of the temperatures 
to obtain PMFs.

\subsubsection{Effect of temperature}

Figure~\ref{pmfcb} shows a representative set of PMFs for different temperatures
(see legend) between three possible pairs of particles (A-A, B-B, and A-B). 
The symbols represent data from the REMD simulations sampled at a single temperature, and the lines are the 
WHAM reconstruction based on the data from all temperatures. The agreement between the 
two estimates clearly indicates that the simulation times are long enough to obtain  
PMFs. 
The interactions between like pairs (A-A and B-B) are purely repulsive at all of the 
temperatures, as shown in Fig.~\ref{pmfcb}b, as these ssDNA sequences were designed to have 
minimal self-interactions~\cite{}. 
Moreover, the repulsive interactions between like pairs are independent of the temperature, 
except at very low pair distances. 

The PMFs between unlike particle pairs (A-B), which are coated with complementary ssDNA with a long
sticky end, are shown in Fig.~\ref{pmfcb}a. 
The free energy as a function of pair distance shows a distinct minimum 
at low temperatures, which corresponds to the bound configuration between particles A and B (stabilized by DNA hybridization). 
Specifically, the general shape of the pair potential and its dependence on temperature (compare Fig. 4a with Fig. 1E from Ref. 46) resemble the experimentally measured effective pair potentials~\cite{Rogers2011}.”
For two particle simulations in a large enough box (approaching zero density 
limit), the minimum in free energy is an important parameter to characterize interactions between particles as a function of system 
parameters such as temperature, DNA grafting density, and length of sticky or spacer end~\cite{Dreyfus2010}. 

Figure~\ref{binding} shows the minimum interaction free energy as a function of temperature, which we find to vary in a 
highly non-linear manner. Previously, Dreyfus {\em et al.}~\cite{Dreyfus2010} also predicted a similar 
non-linear dependence of the minimum interaction free energy on temperature with the help of a 
theoretical model. In their model, the minimum interaction free energy depends on the number of 
bonds formed between the two particles as well as the entropy associated with the selection of 
various combinations of strands or bonds on the two particles; the sharp decrease in the minimum 
interaction free energy as a function of decreasing temperature is most likely due to enhanced 
hydrogen bonding (due to the increase in the number of hydrogen bonds as the temperature decreases and the increased stability of a hydrogen bond with decreasing temperature), which is further facilitated by shorter distances at which the minimum occurs. 
However, in their model DNA is simplified as a rigid rod with a sticky point at the 
end (to represent complementary ssDNA),   
and the two particles are approximated as flat plates. 
In our model, ssDNA strands are relatively flexible, and the length of the sticky end is 
comparable to the length of the spacer. 
Given these differences between our simulation model and the theoretical model  
of  Dreyfus {\em et al.}, the non-linear dependence of the minimum interaction 
free energy on temperature is expected to be a general feature of DNA-mediated particle 
interactions. At low temperatures, the minimum interaction free energy can be quite significant compared to the 
thermal energy, thereby leading to irreversible particle binding.  
Appearance of such irreversible particle binding at temperatures not too far from the desired 
temperature 
range (weak binding regime) due to non-linearity is likely to be an inherent hindrance to particle 
rearrangement, and, therefore, crystallization in DNA-mediated particle assembly. 

\begin{figure}
\includegraphics[width= 0.75\columnwidth]{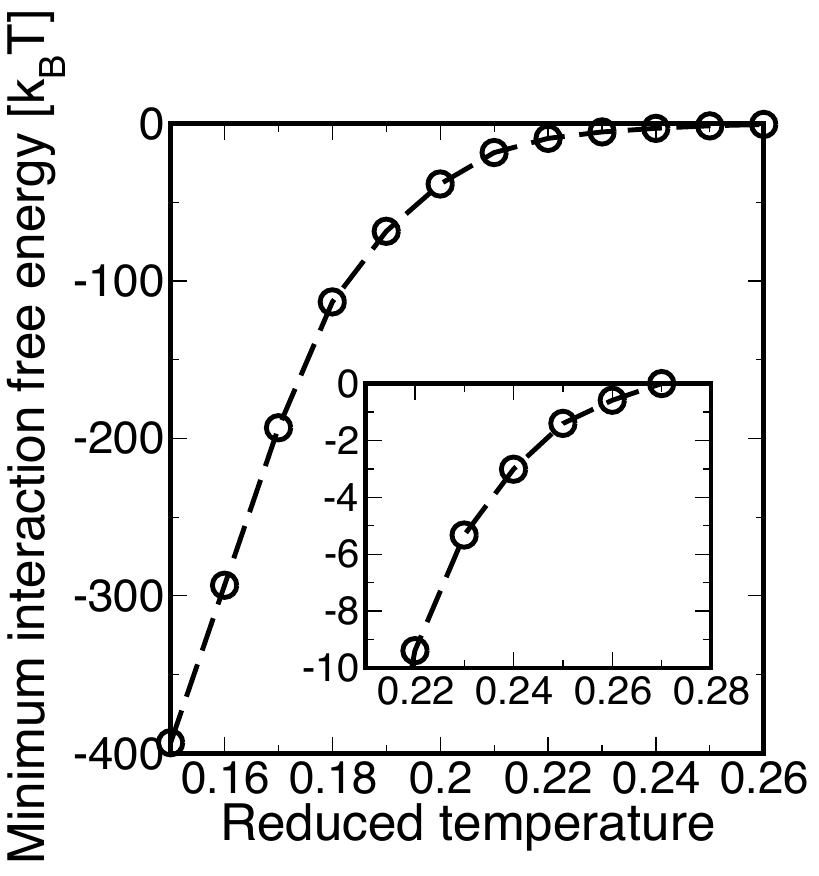}
\caption{The minimum interaction free energy is shown as a function of 
temperature corresponding to system parameters used in Fig.~\ref{pmfcb}a. The inset shows a zoomed-in 
view of the main plot.}
\label{binding}
\end{figure}

\subsubsection{Effect of DNA grafting density}

In experiments, DNA grafting density on the particle surface can be controlled by synthesis methods~\cite{Demers2000,Dmytro2007} as well 
as by adding non-hybridizing ssDNA in the solution buffer~\cite{Dreyfus2009}. By controlling the fraction 
of sticky ends that can hybridize with complementary ssDNA on other particles, it was shown that the 
particle dissociation behavior (melting temperature and sharpness) can be controlled 
quantitatively~\cite{Dreyfus2009,Casey2012}. It is still less well understood how this 
particular parameter can affect the pair interaction potential and the associated assembly mechanism. Here, 
we study the first question by calculating PMFs as a function of DNA grafting density ranging from 1 
DNA molecule per particle to 32 DNA molecules per particle.  
  
Figure~\ref{min} shows the minimum interaction free energy as a function of temperature for varying DNA grafting densities 
for the case of a short sticky end. 
For a given temperature, the minimum interaction free energy becomes lower with 
increasing DNA grafting densities as a greater number of complementary DNA molecules can hybridize between 
a pair of particles. 
According to Dreyfus et al.~\cite{Dreyfus2010}, the non-linear dependence of the minimum interaction free energy can 
be categorized into three different regimes. The weak-binding regime, in which the minimum interaction free energy is comparable to the thermal energy, is most relevant for successful DNA-mediated particle assembly in laboratory experiments. In this regime, the minimum interaction free energy is 
expected to be proportional to the average number of hydrogen bonds formed~~\cite{Dreyfus2010}. 
In Fig.~\ref{min}, we limit the range of the minimum interaction free energy from $-10k_{B}T$ to  $-k_{B}T$, which  
approximately corresponds to the so-called weak-binding regime. We find that even in this regime, the variation in 
minimum interaction free energy as a function of temperature is quite non-linear. 
Similar behavior in the weak 
binding regime was also observed previously by Mirjam {\em et al.}~\cite{Mirjam2011}.

\begin{figure}
\includegraphics[width= 0.9\columnwidth]{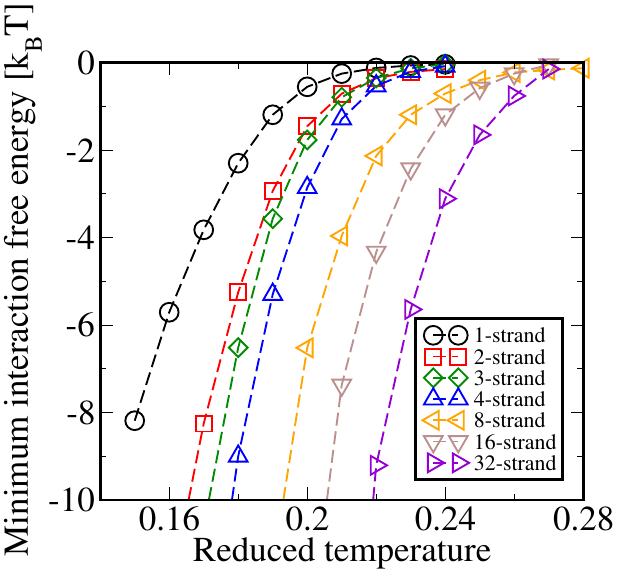}\\
\caption{The minimum interaction free energy is shown as a function of 
temperature for different grafting densities (see legend) for the short sticky end.}
\label{min}
\end{figure}

To identify changes in pair PMF as a function of DNA grafting density, we use two special cases -- constant 
temperature and constant minimum interaction free energy. Figure~\ref{pmfshort}a shows PMFs between unlike particles 
(A-B) for different grafting densities, but the minimum interaction free energy is kept constant at $-4 k_{\mathrm {B}}T$ by changing the 
temperature (see legend). We find that the change in temperature is related logarithmically to the 
DNA grafting density in order to obtain a constant minimum interaction free energy (data not shown). 
Though not entirely unexpected, the DNA grafting density significantly 
alters the shape of the pair interaction potential or PMF. 
The minimum in free energy is quite broad for lower grafting densities, but the attractive well becomes 
narrower with increasing DNA grafting density. This narrowing of the attractive well is most likely due to 
stronger repulsion caused by overlap between grafted DNA molecules at small pair distances.  
Although some previous work has been done relating shape of the pair potential with 
particle dynamics~\cite{Zane2011}, there is no general framework to determine which of these potential shapes 
will be most desirable for ordered particle assembly. Our hypothesis is that intermediate DNA grafting 
densities will yield an optimum in the width of the attractive well so that pair distance can change 
as needed for structural rearrangements, while the resulting ordered structure is still stable without 
significant fluctuations in lattice parameters. 

In Figure~\ref{pmfshort}b, we compare PMFs at the same temperature for varying DNA grafting densities. 
The minimum interaction free energy decreases with increasing DNA grafting density as expected due to the higher 
number of complementary strands that can hybridize. However, the maximum number of hydrogen bonds, defined as the number of hydrogen bonds formed between the two particles at the lowest temperature of our simulation, does not linearly increase with the number of grafted DNA chains (see SM Figure S4~\cite{sm}). This is most likely due to the excluded volume (repulsive) interactions between the DNA molecules on the same particle, thereby limiting the number of strands that can hybridize with increasing DNA grafting density. Similar to Figure~\ref{pmfshort}a, we also observe narrowing of the attractive well in 
addition to deepening at higher DNA grafting densities, again due to enhanced repulsion between 
DNA molecules at small pair distances. 

Finally, we note that the current model, due to its approximate nature, can only provide qualitative changes in the potential function, and therefore the observed changes in the potential width should be interpreted accordingly.

\begin{figure}
\includegraphics[width= 0.9\columnwidth]{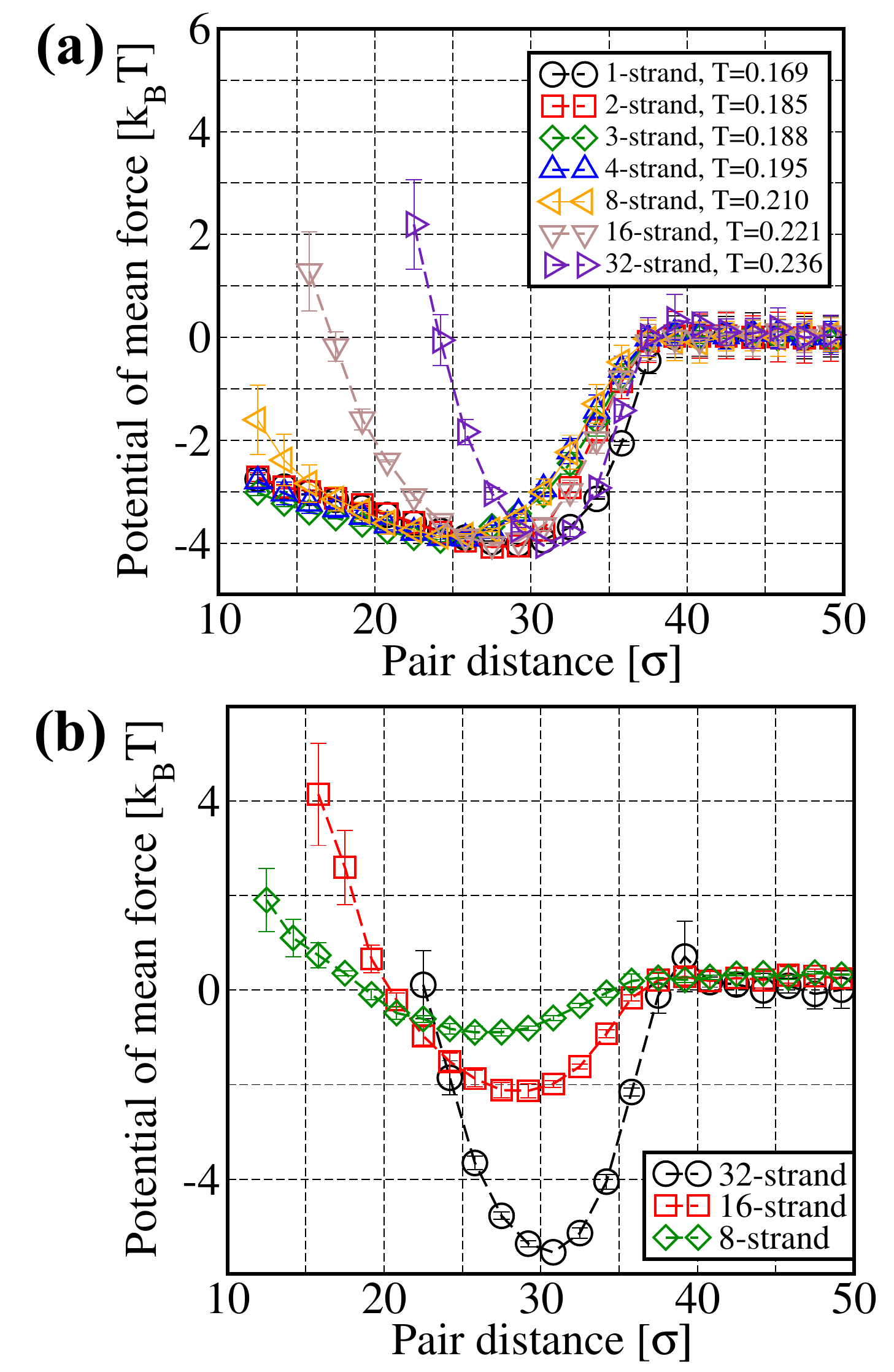}
\caption{(a) Potential of mean force (PMF) for different DNA grafting densities for the short sticky end with the same attraction 
free energy of $-4 k_{B}T$. (b) PMF for different grafting densities for the short sticky end at the same temperature  of 0.23.}
\label{pmfshort}
\end{figure}

\subsubsection{Effect of spacer length}

Experimentally, Jin {\em et al.}~\cite{Jin2003} showed that an increase in the spacer length (the non-hybridizing part 
of the ssDNA) increases the melting temperature of DNA grafted on nanoparticles. Later, 
Sun {\em et al.}~\cite{Sun2005} also observed a similar increase in DNA melting temperature with increasing 
spacer length. 
From molecular simulations of the CG DFP model, Fernando {\em et al.}~\cite{Vargas2011} found that longer spacer
length can actually destabilize a preformed crystal lattice of DFPs at lower temperatures. 
Although there are several studies on the relationship between spacer length and DNA melting temperature, 
relatively little information is available on the dependence of the pair interaction potential between DFPs on spacer length. 

Figure~\ref{spacer} shows PMFs between a pair of unlike DFPs (A-B) when each particle is grafted with 
16 complementary ssDNA molecules with the long sticky end at the same temperature. The minimum interaction free energy decreases slightly (by about 1 $k_{\mathrm B}T$) with increasing spacer lengths 
from 0 to 8 bases.  
This suggests that increasing the spacer length may help in the formation of more hydrogen bonds between 
DFPs, but other parameters such as DNA grafting density and particle curvature may be more important 
determinants of enhanced hydrogen bond formation. 
More importantly, the pair distance at which the minimum in the PMF occurs shifts to 
a higher value with increasing spacer length. These results are consistent with experimental observations, 
as the decrease in the minimum interaction free energy will lead to a higher melting temperature and the increase in 
preferred pair distance will lead to a longer translational lattice parameter~\cite{Macfarlane2011}. 
As the width of the PMF also increases with increasing spacer length, one may expect destabilization of 
certain crystal lattices as observed by Fernando {\em et al.}~\cite{Vargas2011} in their simulations.

\begin{figure}
\includegraphics[width= 0.9\columnwidth]{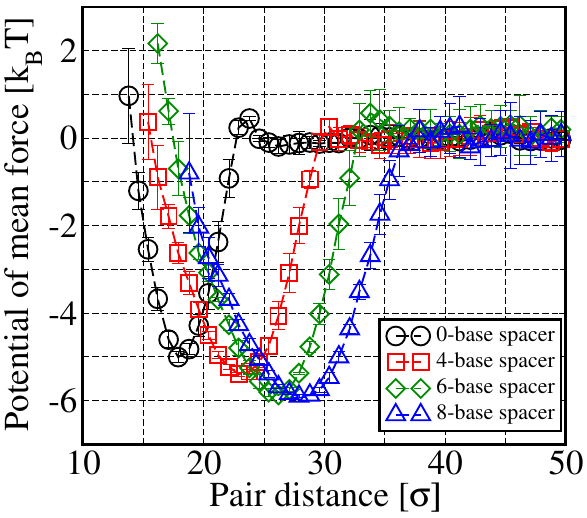}
\caption{Effect of spacer length on the potential of mean force for long sticky end with 16 DNA molecules per particle at temperature 0.224.} 
\label{spacer}
\end{figure}

\subsubsection{Effect of sticky end length}

\begin{figure}
\includegraphics[width= 0.9\columnwidth]{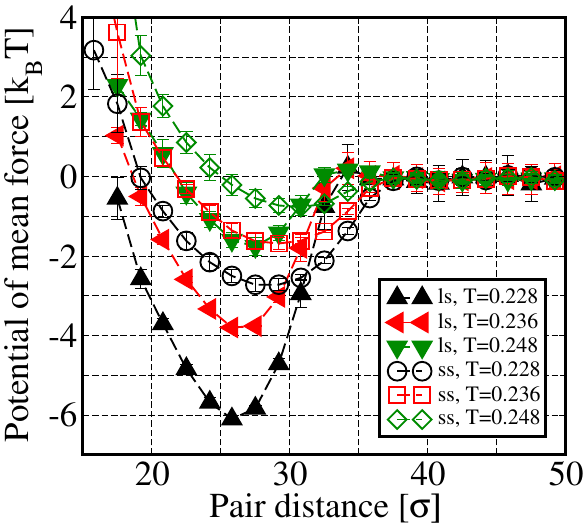}
\caption{Comparison of potential of mean force between short and long sticky ends at three different 
temperatures. ss: short sticky end; ls: long sticky end.}
\label{stickyend}
\end{figure}

In addition to varying the DNA grafting density, one can also alter the minimum interaction free energy 
between DFPs by changing the number of complementary base pairs (or type of base pairs formed,   
A:T versus G:C~\cite{Seifpour2013}) of the sticky end. Crocker {\em et al.}~\cite{Milam2003} 
found that reducing the sticky end length leads to a decrease in melting temperature. Similar conclusions 
were also drawn by others using separate linker-mediated interactions between DFPs~\cite{Harris2005}. 
Figure~\ref{stickyend} compares PMFs obtained for a pair of DFPs with short and long sticky ends for three  
temperatures. Both of the DNA sequences (long and short sticky ends) have a total of 18 bases, out of 
which 
7 and 12 are part of the sticky end for short and long sticky ends, respectively. 
Intuitively, varying the length of the sticky end (while keeping the total DNA length constant) can have two 
effects on the shape of the PMF -- change in the depth and change in the width of the attractive potential well. As shown in 
Fig.~\ref{stickyend}, the minimum interaction free energy for the short sticky end is higher than for the long sticky end 
at the same temperature, as expected. The minimum in free energy for the short sticky end is also shifted 
to longer pair distances due to shorter overlap between shorter sticky ends as compared to long sticky ends. Interestingly, the PMF well in the case of the short sticky end is wider as
compared to that of the long sticky end. 
Current CG DFP models, which assume a point-like sticky end, cannot capture the effect of sticky end length on 
the shape of the PMF. This is not expected to be limiting for small variations in sticky end length but 
may be important for large changes in the sticky end length to model associated changes in the particle 
assembly behavior. 

\section{conclusions}

In this paper, we present a new coarse-grained model for studying the behavior of DNA-functionalized 
particles. The coarse-grained DNA model used here provides explicit DNA representation (at the nucleotide level) and complementary interactions between Watson-Crick base pairs, which lead to the formation of ssDNA hairpins and dsDNA. Aggregation between multiple complementary strands is prevented. We use this model to calculate the dependence of the free energy on the distance between a pair of DNA-coated particles as a function of temperature, DNA grafting density, and lengths of sticky end and spacer DNA. 
The change in the minimum interaction free energy as a function of system temperature is 
found to be non-linear even in the weak-binding regime. 
Our results of particle pair potentials as a function of system parameters can also be useful for future 
design of crystal lattices based on DNA-mediated particle assembly. For example, 
Macfarlane {\em et al.}~\cite{Macfarlane2011}, suggests that the lattice spacing can be adjusted by 
varying the hydrodynamic size ratio between two types of particles grafted with complementary DNA 
molecules. Based on our effective pair potentials, one should be able to modify the lattice spacing 
with the same hydrodynamic particle size ratio by varying the sticky end length while keeping the total 
DNA length the same. Moreover, we find that several system parameters (such as DNA grafting density and sticky end 
length) can be used to change the width of the attractive well.
We hypothesize that shallower and broader attractive potentials may be preferable for particle assembly 
in ordered structures due to dynamic DNA hybridization~\cite{Ting2013}, but may also lead to lower lattice 
stability. In the future, we plan to address these questions with multiparticle assembly simulations using the new 
coarse-grained model. 

\section{Acknowledgments}
We thank Drs. Hasan Zerze and Young Chan Kim (Naval Research Laboratory) for helpful discussions. 
This work was supported by the National Science Foundation grant CBET-1120399.  Use of the high-performance computing capabilities of the Extreme Science and Engineering Discovery Environment (XSEDE), which is supported by the National Science Foundation grant no. TG-MCB-120014, is also gratefully acknowledged.
 
\bibliographystyle{apsrev4-1}

%

\end{document}